\documentclass[aasmacros]{aa}

\usepackage{amsmath}
\usepackage{wasysym}
\usepackage{marvosym}
\usepackage{txfonts}
\usepackage{placeins}
\usepackage{siunitx}
\usepackage{enumitem}
\DeclareSIUnit[]\astronomicalunit{\text{au}}

\usepackage{natbib}
\bibpunct{(}{)}{;}{a}{}{,} 

\usepackage{graphicx}
\usepackage[dvipsnames,table]{xcolor}
\usepackage{subcaption}

\usepackage{scalerel}
\usepackage{tikz}
\usetikzlibrary{svg.path}

\definecolor{orcidlogocol}{HTML}{A6CE39}
\tikzset{
  orcidlogo/.pic={
    \fill[orcidlogocol] svg{M256,128c0,70.7-57.3,128-128,128C57.3,256,0,198.7,0,128C0,57.3,57.3,0,128,0C198.7,0,256,57.3,256,128z};
    \fill[white] svg{M86.3,186.2H70.9V79.1h15.4v48.4V186.2z}
                 svg{M108.9,79.1h41.6c39.6,0,57,28.3,57,53.6c0,27.5-21.5,53.6-56.8,53.6h-41.8V79.1z M124.3,172.4h24.5c34.9,0,42.9-26.5,42.9-39.7c0-21.5-13.7-39.7-43.7-39.7h-23.7V172.4z}
                 svg{M88.7,56.8c0,5.5-4.5,10.1-10.1,10.1c-5.6,0-10.1-4.6-10.1-10.1c0-5.6,4.5-10.1,10.1-10.1C84.2,46.7,88.7,51.3,88.7,56.8z};
  }
}

\newcommand\orcidicon[1]{\href{https://orcid.org/#1}{\mbox{\scalerel*{
\begin{tikzpicture}[yscale=-1,transform shape]
\pic{orcidlogo};
\end{tikzpicture}
}{|}}}}

\usepackage{hyperref}
\hypersetup{draft=False, colorlinks=true, citecolor=blue, urlcolor=blue, linkcolor=red, linktoc=all}

\usepackage{pifont}

\def\msun{M_\odot}

\def\f1{f_{\rm I}}

\def\beq{\begin{equation}}
\def\eeq{\end{equation}}

\def\t2{\tau_{\rm II}}

\def\sigmas0{\Sigma_{\rm s,0}}

\def\s0{S_0}

\def\apjl{ApJ}                
\def\apjs{ApJS}               
\def\ao{Appl.Optics}          
\def\aap{A\&A}                

\def\({\left(}
\def\){\right)}
\def\<{\left<}
\def\>{\right>}

\defcitealias{2021A&A...645A..43S}{Paper~I}
\defcitealias{2018MNRAS.475.5618B}{B18}
\defcitealias{Kimura2016}{K16}
\defcitealias{2017MNRAS.469.3813H}{HK17}
\defcitealias{2008ApJ...685..560D}{DL08}
\defcitealias{2010ApJ...724..730D}{DL10}
\defcitealias{2019MNRAS.486.4398F}{FNS19}
\defcitealias{1986ApJ...309..846L}{LP86}
\defcitealias{2002MNRAS.334..248A}{A02}
\defcitealias{2015A&A...582A.112B}{BLJ15}

\begin{document}




\title{The link between infall location, early disc size, and the fraction of self-gravitationally fragmenting discs}

   \author{O. Schib\inst{\ref{bern},\ref{uzh}}
          \and
          C.~Mordasini\inst{\ref{bern}}
          \and
          R.~Helled\inst{\ref{uzh}}
          }

   \institute{
Physics Institute,
University of Bern, Gesellschaftsstrasse 6, 3012 Bern, Switzerland, 
\email{oliver.schib@space.unibe.ch}
\label{bern}
\and
Institute for Computational Science,
Universit\"at Z\"urich, Winterthurerstrasse~190, 8057~Z\"urich, Switzerland
\label{uzh}
             }

\date{Received August 22, 2022 / Accepted November 10, 2022}

\abstract
{Many protoplanetary discs are self-gravitating early in their lives. If they fragment under their own gravity, they form bound gaseous clumps which may evolve to become giant planets. Today, the fraction of discs that undergo fragmentation, and therefore also the frequency of conditions that may lead to giant planet formation via gravitational instability,  is still  unknown.}
{We study the formation and evolution of a large number of star-disc systems focusing on the discs' early sizes and their likelihood to fragment. We investigate how the fraction of discs that fragments depends on the disc size distribution at early times.}
{We perform a population synthesis of discs from formation to dispersal. In varying the infall radius, we study the relationship of the early disc size with fragmentation. Furthermore, we investigate how stellar accretion heating affects the fragmentation fraction.}
{We find that discs fragment only if they become sufficiently large early in their lives. This size depends sensitively on where mass is added to the discs during the collapse of their parent molecular cloud core. Infall locations derived from pure hydrodynamic (non-ideal magnetized collapse) simulations lead to large (small) discs and a 22\% (0\%) fragmentation fraction in populations representative of the initial mass function. However, the resulting synthetic disc size distribution is larger (smaller) than the observed Class 0 disc size distribution. By choosing intermediate infall locations leading to a synthetic disc size distribution that is in agreement with the observed one, we  find a fragmentation fraction between 0.1 and 11\%, depending on the efficiency of stellar accretion heating of the discs.}
{We conclude that the frequency of fragmentation is strongly affected by the early formation process of the disc and its interaction with the star. The early disc size is mainly determined by the infall location during the collapse of the molecular cloud core and controls the population-wide  frequency of fragmentation. Stellar accretion heating also plays an important role for fragmentation and must be studied further. Our work is an observationally-informed step towards a prediction of the frequency of giant planet formation by gravitational instability. Upcoming observations and theoretical studies will deepen our understanding of the formation and early evolution of discs in the near future. This will eventually allow to understand how infall, disc morphology, giant planet formation via gravitational instability, and the observed extrasolar planet population are linked. }


\maketitle

\section{Introduction}\label{sect:Introduction}
Protoplanetary discs are of central importance in astrophysics: they are the birth places of planets.
Two main pathways for planet formation exist: core accretion \citep{safronov1972,Pollack1996,2004ApJ...604..388I,mordasinialibert2012a} and gravitational instability (GI, \citealp{1951PNAS...37....1K,Cameron1978,1997Sci...276.1836B,2016ARA&A..54..271K}).
GI remains the leading formation mechanism for some observed giant planets on wide orbits \citep{2008Sci...322.1348M,2010Natur.468.1080M,2018ApJ...860L..12T} and may explain the formation of giants around low-mass stars \citep{2019Sci...365.1441M}.
GI has also been proposed for the formation of intermediate-mass planets \citep{2021NatAs.tmp...26D} and very young giants \citep{2022NatAs.tmp...76C,2021MNRAS.504.2877C}.
Fragmentation, the collapse of parts of a protoplanetary disc to form a bound object, is a necessary condition for GI.
It is still debated Whether a bound clump formed by GI can survive to become a giant planet.
The interplay among several physical processes like migration, accretion of gas and solids, dynamical collapse, grain sedimentation and core formation needs to be studied in order to answer this question  \citep{Bodenheimer1980,2008Icar..195..863H,2008Icar..198..156H,2013MNRAS.432.3168F,2017PASA...34....2N,2013MNRAS.432.3168F}.

While many questions around GI are still open, the first step of fragmentation has been studied extensively using 2D and 3D hydrodynamic simulations \citep{1964ApJ...139.1217T,2001ApJ...553..174G,2002Sci...298.1756M,2010ApJ...719.1896V,2017ApJ...848...40B,2017ApJ...847...43D,2020ApJ...904...55J}.
Discs are found to be prone to fragmentation in the outer regions (tens of $\SI{}{au}$) if they are cold and massive enough \citep{1998ApJ...503..923B,2000ApJ...536L.101B,2001ApJ...553..174G,Clarke2009a,2004MNRAS.355..543R,2005MNRAS.364L..56R}.
However, while many discs are expected to undergo a phase of gravitational instability early in their evolution,  \citep{Durisen2006} it is uncertain how many of them  fragment. 
Fragmentation could also be triggered externally by binary companion in discs that are not massive enough to fragment on their own \citep{2022MNRAS.511..457C}.
Gravitational instability also drives transport of angular momentum via spiral arms and discs can remain in a quasi-steady state of self-regulation \citep{2004MNRAS.351..630L,Cossins2009,Rice2016}.
A better understanding of the frequency of fragmentation is important for population synthesis studies of GI as it determines the expected number of clumps. The number of clumps is in turn an upper bound for the number of planets that may from via gravitational instability. A theoretical prediction of the fragmentation fraction is thus a necessary step towards a  prediction of the frequency of planet formation via GI. This is of high interest for comparisons with the extrasolar planet population.  

Even if GI does not produce a significant number of planets, the formed clumps may still be responsible for free-floating planetary mass objects \citep{2018MNRAS.474.5036F} or episodic accretion \citep{2014prpl.conf..387A}.

In previous work (\citealp{2021A&A...645A..43S}, hereafter Paper~I) we performed a population synthesis of protoplanetary discs with a focus on fragmentation.
We included the formation of the star-and-disc system by infall from the molecular cloud core (MCC), since fragmentation is most likely to occur during or shortly after disc formation where the disc-to-star mass ratio is highest.
It was found that fragmentation depends sensitively on the disc formation process.
If young discs are compact, as expected by some non-ideal magnetohydrodynamic (MHD) simulations of cloud collapse \citep{Hennebelle2016}, fragmentation is suppressed completely.
If discs are large already early on, as expected from radiation hydrodynamic simulations \citep{2018MNRAS.475.5618B}, fragmentation may be prevalent.
Our comparison with observed sizes of Class~0 discs suggested that real disc sizes lie somewhere between the two extremes. 
This raises the question of the likelihood of discs to fragment if they have sizes consistent with observations.
While we studied a broad parameter space in \citetalias{2021A&A...645A..43S}, in this paper we focus on a specific aspect: how the disc size distribution from infall sets the fraction of self-gravitationally fragmenting discs.

First, we briefly review the model. Then we describe our analysis and present the results. A short discussion is followed by the summary and conclusions.

\section{Model and investigated parameter space}\label{sect:Model}
Our model is described in detail in \citetalias{2021A&A...645A..43S}.
We apply it without changes. 
The most important aspects of the model are summarised here for convenience.

The model consists of a 1D, vertically integrated gas disc around a single star.
We use cylindrical polar coordinates, with $r$ denoting the radial direction.
The disc's mid-plane is located at $z=\SI{0}{au}$.
The surface density $\Sigma$ evolves in time $t$ according to the viscous evolution equation (\citealp{1952ZNatA...7...87L} and \citealp{1974MNRAS.168..603L}):
\begin{equation}
\label{evogeneral}
	\frac{\partial \Sigma}{\partial t} = \frac{3}{r}\frac{\partial}{\partial r}\left[ r^{1/2} \frac{\partial}{\partial r}\left(\nu \Sigma r^{1/2}\right)\right] + S.
\end{equation}
$S \equiv S(r,t)$ is a source/sink term:
\begin{equation}
    S(r,t) = S_\mathrm{inf}(r,t) - S_\mathrm{int}(r,t) - S_\mathrm{ext}(r,t) - S_\mathrm{frag}(r,t).
\end{equation}
The summands describe, from left to right, the infall from the MCC, internal \Citep{2001MNRAS.328..485C} and external \Citep{2003ApJ...582..893M} photoevaporation, and the removal of mass due to fragmentation.
Detailed descriptions and analytical expressions are given in \citetalias{2021A&A...645A..43S}.
We assume a short phase of infall at the beginning of our simulations.
The precise implications of this assumption are difficult to assess. We discuss this further in Sect.~\ref{sect:Discussion}.
The constant infall rate is chosen based on a selection of systems from the radiation hydrodynamic simulation of star formation in \citet{2018MNRAS.475.5618B} (B18).

Our calculation of the viscosity is based on the $\alpha$-prescription \Citep{1973A&A....24..337S}: $\nu = \alpha c_\mathrm{s^2}/\Omega$, with $c_\mathrm{s}$ the isothermal sound speed and $\Omega$ the disc's angular frequency.
We use different values for $\alpha$ to describe different physical processes. During the infall phase, the disc-to-star mass ratio can reach values of \num{0.5} or higher, and the transport of angular momentum is dominated by gravitational torques \citep{shutremaine1990,2011MNRAS.413..423H}. We use the prescription given in Eq.~(32) of \citet{2010ApJ...708.1585K} to account for these effects. After the infall phase, the transport of angular momentum can remain very high as long as the disc is self-gravitating \citep{2008ApJ...681..375K}. We use the corresponding parametrisation for the transport through spiral arms given in \citet{2010ApJ...713.1143Z} in this case.
In the absence of gravitational torques we apply a minimal background value of \num{0.01} \citep{2016MNRAS.461.2257K} that describes any other physical process responsible for the transport of angular momentum, like the magneto-rotational instability (MRI, \citealt{1991ApJ...376..214B}), hydrodynamic instabilities or MHD winds \citep{2014prpl.conf..411T}.
We note that MRI is no longer considered the main source of turbulence in protoplanetary discs due to their weak ionisation (e.g. \citealt{2022arXiv220309821L,2021MNRAS.507.1106C}).
If the long-term evolution is indeed better described by a value of $\alpha$ between \num{e-4} and \num{e-3}, our model would predict disc lifetimes much longer than observed, as discussed in \citetalias{2021A&A...645A..43S}, unless photoevaporation is much stronger than we assume.

The disc's midplane temperature is calculated assuming an energy balance on the surface.
The following processes are considered: viscous heating, shock heating from the infall, irradiation from a constant background and irradiation by the star.
The irradiation temperature is calculated as \citep{2005A&A...442..703H}:
\begin{equation}\label{eq_irr}
	T_\mathrm{irr} = T_* \left[ \frac{2}{3 \pi} \left( \frac{R_*}{r} \right)^3 + \frac{1}{2} \left( \frac{R_*}{r} \right)^2 \left(\frac{\mathrm{d} \ln(H)}{\mathrm{d} ln(r)}-1 \right) \right]^{1/4}.
\end{equation}
In Eq.~\ref{eq_irr}, $T_*$ and $R_*$ are the star's temperature and radius, respectively.
We set $\mathrm{d} \ln(H)/\mathrm{d} \ln(r) = 9/7$ for numerical reasons \citep{2005A&A...442..703H}.
The star's intrinsic temperature and radius are interpolated from stellar evolution tables \citep{2008ASPC..387..189Y}.
As the disc evolves, some of its mass crosses the inner truncation radius assumed to be constant at $\SI{0.05}{au}$.
It is accreted onto the central star.
For the majority of the simulations, we assume that half of the gravitational binding energy is radiated away, thermalised and heating the disc.
For the total stellar luminosity $L_*$ we write \Citep{2009ApJ...702L..27B,2018A&A...618A...7V,2020A&A...644A..41O}:
\begin{equation}\label{eq_lstar}
	L_* = L_\mathrm{int} + f_{\rm acc}\frac{G M_* \dot{M_*}}{2 R_*},
\end{equation}
where $L_\mathrm{int}$ is the star's intrinsic luminosity and the second term is half the stellar accretion (shock) luminosity, with $f_{\rm acc}$ an efficiency factor, $G$ the gravitational constant, $M_*$ the stellar mass and $\dot{M_*}$ the accretion rate of disc material onto the star. 
The effective stellar temperature in Eq.~\ref{eq_irr} is then calculated as:
\begin{equation}
	T_\mathrm{*} = \left( \frac{L_*}{4 \pi R_*^2 \sigma} \right)^{1/4}
\end{equation}
with $\sigma$ the Stefan-Boltzmann constant.

Nominally, we assume, $f_{\rm acc}$=1 \citep{2009ApJ...702L..27B,2016MNRAS.461.2257K,2020A&A...644A..41O,2018MNRAS.475.2642K}.
Treating stellar irradiation this way is a strong assumption and likely an overestimate, since the accretion luminosity is absorbed/thermalised in the infalling material.
We discuss this further in Sect.~\ref{sect:Discussion}.
There is also some shock heating related to material from the MCC as it reaches the disc.
\citetalias{2021A&A...645A..43S} suggest that its effect is small compared to the accretion on the star and we include it in all runs.

During the disc's evolution, the criteria for fragmentation are checked in every time step.
We use the same criteria as in \citetalias{2021A&A...645A..43S}. The main (necessary) condition is the Toomre criterion \Citep{1964ApJ...139.1217T}:
\begin{equation}\label{eq:toomre}
  Q_\mathrm{Toomre} = \frac{c_s \kappa}{\pi G \Sigma} <1,
\end{equation}
where $Q_\mathrm{Toomre}$ is the Toomre parameter and $\kappa$ the epicyclic frequency.
If the disc is supplied by infalling material faster than it can be transported away, it will invariably fragment when $Q_\mathrm{Toomre}$ drops below unity \citep{2009ApJ...695L..53B}. This is called the ``infall dominated regime''. The condition for this regime is given in \citetalias{2021A&A...645A..43S} (Eq.~(25).
If the infall rates are not high enough, the disc transitions into the ``cooling dominated regime''. In that case it can only fragment with sufficient cooling,  which can be stated as $t_\mathrm{cool} < \beta_c$ \citep{2001ApJ...553..174G}, where we set the critical cooling $\beta_c$ to \num{3} \citep{2017ApJ...847...43D}. The cooling timescale $t_\mathrm{cool}$ is given in Eq.~(27) in \citetalias{2021A&A...645A..43S}.

When the disc fragments, a bound clump composed predominantly of gas forms.
Such a clump may undergo a number of processes like migration, accretion and disruption.
The combined outcome of these is uncertain.
In this work we avoid this complexity and follow the approach described in  Sect.~2 of \citetalias{2021A&A...645A..43S}: we assume the clump migrates to the inner disc quickly and accretes on the central star.
The initial fragment mass of
\begin{equation}\label{eq_mtoomre}
    M_F = 1.6 c_s^3/(G \Omega)
\end{equation}
(\citealt{2010Icar..207..509B}, tough see also \citealt{2011MNRAS.417.1928F}) is removed from the disc an added to the star.

In \citetalias{2021A&A...645A..43S}, we performed a disc population synthesis with five runs, each of them consisting of \num{10000} simulations.  
We varied several parameters to investigate how they affect the masses, sizes and fragmentation of the discs.
The run ``hydro''  (the baseline case) was set up with initial conditions extracted from results presented in the hydrodynamic simulation of star formation by \citet{2018MNRAS.475.5618B}.
This includes the infall location, which was chosen to approximately have the same angular momentum of the discs compared to \citet{2018MNRAS.475.5618B}.
Based on the ``hydro'' run, we performed three additional runs where we varied the background viscosity, the accretion heating efficiency and the initial fragment mass.
We also performed a run where we chose the infall location differently. The run ``MHD'' is described in more detail below.
The simulations we conducted in the present work are executed in an analogous fashion as the ones in \citetalias{2021A&A...645A..43S}:
Every simulation starts with a very low stellar mass and disc mass ($\sim \SI{0.01}{\msun}$), is supplied with infalling material for a given length of time and then evolves further until the disc dissolves.
First, we investigated the dependency of the fragmentation fraction on the the early disc size distribution.
For this, we performed a number of disc population syntheses where we vary the infall radii.
A prediction for the early disc size is given in \citet{Hennebelle2016}.
The authors perform 3D non-ideal MHD simulations and show that the disc radii at early stages of disc formation agree within a factor of two with their analytic expression:
\begin{equation}\label{eq_rhen}
    r_\mathrm{H16} = \SI{18}{au} ~ \left( \frac{A}{\SI{0.1}{s}} \right)^{2/9} \left( \frac{B_z}{\SI{0.1}{G}} \right)^{-4/9} \left( \frac{M_{*\mathrm{d}}}{\SI{0.1}{\msun}} \right)^{1/3}.
\end{equation}
In Eq.~\ref{eq_rhen}, $A$ is a measure of the ambipolar diffusivity, $B_z$ denotes the magnetic field in the inner part of the core and $M_{*\mathrm{d}}$ the combined mass of star and disc.
In \citetalias{2021A&A...645A..43S} we performed a run ``MHD'' in which the infalling material was deposited at an infall radius constant in time that lies close to the star (a few au, where the specific value depends on stellar mass). 
We demonstrate that the disc radii at the end of the infall phase agree well with Eq.~\ref{eq_rhen} for $A=\SI{0.1}{s}$ and $B_z=\SI{0.1}{G}$.
We find that none of these discs fragmented due to their small sizes. In \citetalias{2021A&A...645A..43S} we also performed a ``hydro'' run, where the infall radii were taken from the (non-magnetized) radiation-hydrodynamic simulations  of \citetalias{2018MNRAS.475.5618B}. This run produced discs of a much larger size  of which a significant fraction (22\%) fragment\footnote{We note that the fragmentation fractions given in \citetalias{2021A&A...645A..43S} were calculated as the ratio of simulations that fragment to the total number of simulations (\num{10000}). Here we are including the relative likelihoods of different final stellar masses according to the IMF in the calculation. This leads to lower values for the fragmentation fraction (e.g. 22\% instead of 45\% reported in \citetalias{2021A&A...645A..43S}). We do this throughout the present work for consistency.}. Interestingly, the resulting synthetic disc radius distribution in these two runs brackets the observed disc radius distribution  \citep{2020ApJ...890..130T}.

For the present work we performed in total six new runs. In the first three new runs (``5x MHD'', ``8X MHD'' and ``12x MHD'') we subsequently increased the infall radii.
We did this by multiplying the (small) infall radii from the ``MHD'' run from \citetalias{2021A&A...645A..43S} by a constant factor that differs in each of the new runs: 5, 8 and 12.
A different value for the infall radius was used in the ``MHD'' run in each mass bin in order for the early disc radius to agree with Eq.~\ref{eq_rhen}. In all but the least massive system, this radius is close to \SI{2}{au}. Therefore, typical values for the new runs are \num{10}, \num{16} and \SI{24}{au}.

In a second step, we modified the infall radii more finely in order to reproduce observed radii of Class~0 discs (run OBS\_IRR). All these runs assume a full efficiency of stellar accretion heating ($f_{\rm acc}=1).$
In a last step, we studied the effect of accretion heating. We performed two additional runs (``OBS\_NOIRR'' and ``OBS\_REDIRR'' ) with inactive ($f_{\rm acc}$=0) or reduced accretion heating ($f_{\rm acc}$=1/12).
Table~\ref{table_runs} gives an overview of the six runs performed in this work.
\begin{table}
\centering
\begin{tabular}{ccccccc}  
\hline\hline
\begin{tabular}[c]{@{}c@{}}Run\end{tabular} &
\begin{tabular}[c]{@{}c@{}}Infall radius\end{tabular} &
\begin{tabular}[c]{@{}c@{}}Stellar accretion\\ heating efficiency $f_{\rm acc}$\end{tabular} \\
\hline
1 & ``\num{5}$\times$ MHD'' & 1 \\
2 & ``\num{8}$\times$ MHD'' & 1 \\
3 & ``\num{12}$\times$ MHD'' & 1 \\
4 & OBS\_IRR   & 1 \\
5 & OBS\_NOIRR   & 0   \\
6 & OBS\_REDIRR   & 1/12   \\
\hline
\end{tabular}
\caption[]{Overview of the runs. The infall radius is the location at which the infalling material is deposited in the disc (see Sect.~\ref{sect:Model}). In runs OBS\_IRR, OBS\_NOIRR and OBS\_REDIRR, the infall radii were chosen to match observed disc sizes. In run OBS\_REDIRR the accretion heating was reduced by a factor of twelve relative to runs 1-4.}
\label{table_runs}
\end{table}
As in \citetalias{2021A&A...645A..43S}, we chose the duration of the infall (the length of the constant infall phase) in such a way, that the distribution of stellar masses at the end of the simulations agrees with the observed initial mass function (IMF) \citep{2005ASSL..327...41C}. A figure demonstrating this can be found in App.~\ref{app:prop}.
It shows the stellar number density in \SI{}{pc^{-3}} per logarithmic interval of mass for all the runs performed in this work.
The agreement is reasonable for all runs across the parameter space from \SIrange{0.05}{5}{\msun} we studied.
In order to compensate for the decreasing number density towards high masses, we divided this parameter space into \num{100} logarithmically spaced mass bins and set up our simulations in such a way that we obtain approximately \num{100} systems with a final stellar mass lying in each bin.
This ensures our statistics are reasonably robust across the entire mass range.

\section{Results}\label{sect:result}
In this section, we present the results of our simulations.
First, we discuss how different infall radii influence the early disc sizes, then we concentrate on the influence of stellar accretion heating on discs with similar early disc sizes.

An overview of the results for all six runs is given in Table~\ref{table_results}.
The numbers shown represent weighted mean values calculated assuming that the final distribution of stellar masses agrees with the IMF.
The disc sizes are increasing from run ``5x~MHD'' to ``12x~MHD'' as expected, while they are the same for the remaining runs as further discussed in Sect.~\ref{subs:radii}.
The fragmentation fractions change significantly for different runs as discussed in Sect.~\ref{subs:radii} to \ref{subs:irr}.
On the contrary, disc masses and lifetimes are very similar.
We discuss these in Appendix~\ref{app:prop}.
\begin{table}
  \centering
  \begin{tabular}{ccccc}  
  \hline\hline
  \begin{tabular}[c]{@{}c@{}}Run\end{tabular} &
  \begin{tabular}[c]{@{}c@{}}$R_\mathrm{disc,infall}$\\$(\mathrm{au})$\end{tabular} &
  \begin{tabular}[c]{@{}c@{}}$M_\mathrm{disc,infall}$\\$(\msun)$\end{tabular} &
  \begin{tabular}[c]{@{}c@{}}fragmentation\\fraction\\$(\%)$\end{tabular} &
  \begin{tabular}[c]{@{}c@{}}$t_\mathrm{NIR}$\\$(\mathrm{Myr})$\end{tabular} \\
  \hline
  1 &  $\num{60(10)}$ &  $\num{0.16(2)}$ & 0  & $\num{4.2(1)}$ \\
  2 &  $\num{76(20)}$ &  $\num{0.18(2)}$ & 0.19 & $\num{4.3(1)}$ \\
  3 &  $\num{110(30)}$ &  $\num{0.22(2)}$ & 3.5  & $\num{4.5(1)}$\\
  4 &  $\num{66(20)}$ &  $\num{0.18(2)}$ & 0.12 & $\num{4.4(1)}$\\
  5 &  $\num{66(10)}$ &  $\num{0.15(2)}$ & 25   & $\num{4.4(1)}$\\
  6 &  $\num{66(10)}$ &  $\num{0.16(2)}$ & 11   & $\num{4.4(1)}$\\
  \hline
  \end{tabular}
  \caption[]{An overview of the results. Listed are the disc's radii and masses at the end of the infall phase, the fraction of discs that fragment, and the reduced disc lifetimes (see Sect.~\ref{sect:result} and Appendix~\ref{app:prop} for further details).}
  \label{table_results}
\end{table}

\subsection{Infall radii and early disc radii}\label{subs:radii}

\begin{figure}
  \includegraphics[width=\linewidth]{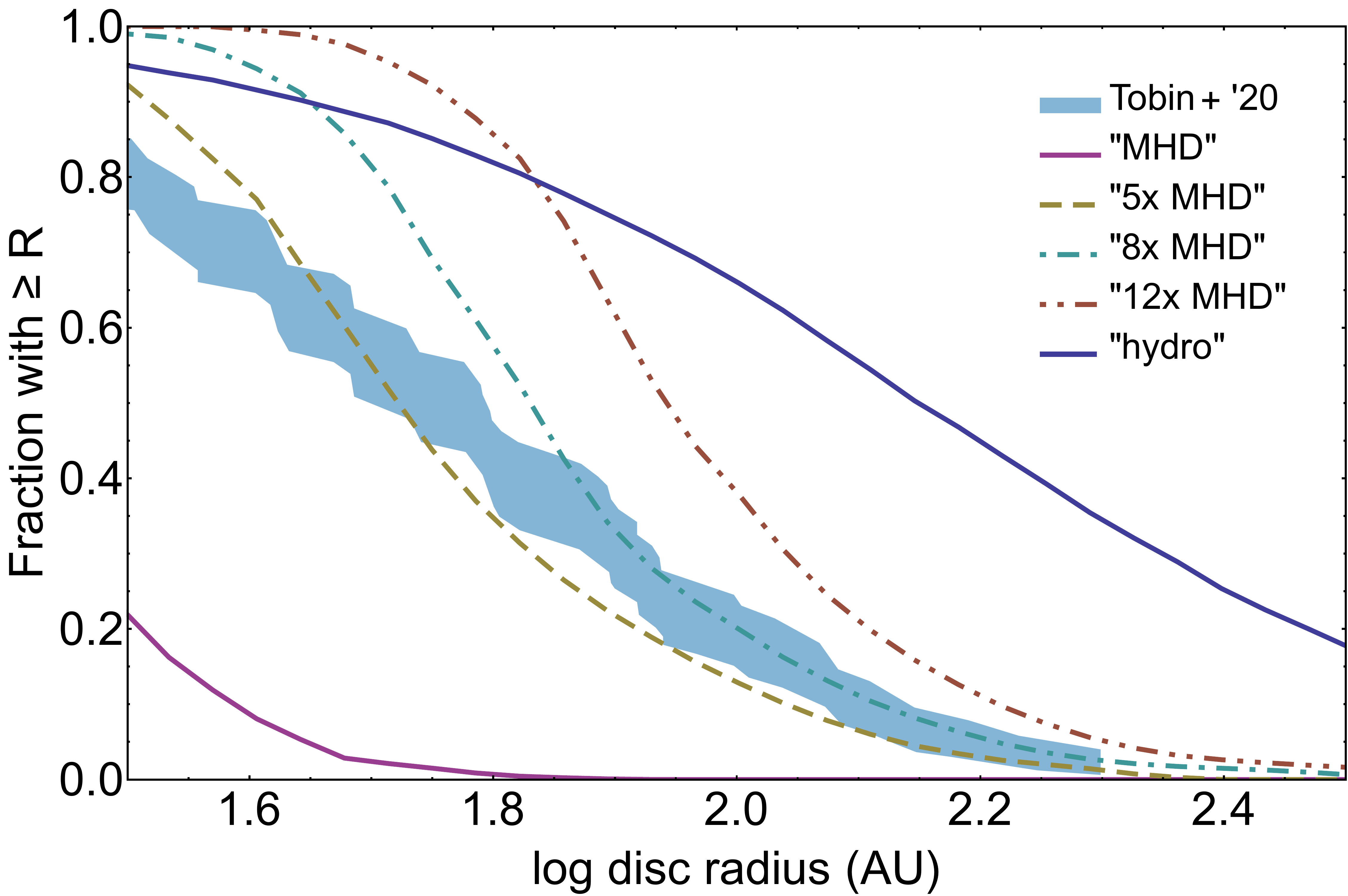}
  \caption{Cumulative distributions of the disc radii at the end of the infall phase for runs ``5x~MHD'', ``8x~MHD'' and ``12x~MHD''. The runs ``MHD'' and ``hydro'' from \citetalias{2021A&A...645A..43S} are shown for reference. The blue shaded region shows the sizes of observed Class~0 discs for systems in isolation from \citet{2020ApJ...890..130T}.}
  \label{fig_rad}
\end{figure}

Figure~\ref{fig_rad} depicts the disc radii of the first three runs, measured at the end of the infall phase, as cumulative distributions.
The disc radius is defined as the radius containing \SI{63.2}{\%} of the disc's mass \citepalias{2018MNRAS.475.5618B}.
The figure also shows the observational Result from \citet{2020ApJ...890..130T}.
The authors perform a multi-wavelength survey of hundreds of protostars and use dust continuum emission to measure Class 0 dust disc radii.
During this early phase the gas disc radii should not yet differ very much from those obtained from continuum emission (\citealt{2014ApJ...780..153B}, \citetalias{2018MNRAS.475.5618B}).
Nevertheless, this is a clear limitation of our study and we discuss this further in Sect.~\ref{sect:Discussion}.

Fig.~\ref{fig_rad} shows that runs 1 to 3 cover a region of parameter space much closer to observed disc radii than the ``hydro'' and ``MHD'' runs we performed in \citetalias{2021A&A...645A..43S}.
It is therefore interesting to check which of these systems fragment. 
The region in a disc prone to fragmentation is typically outside of $\sim \SI{50}{au}$.
There, run ``5x~MHD'' has radii at the lower end of the observed population, 
run ``8x~MHD'' is roughly in agreement,  and run ``12x~MHD'' exhibits larger radii. This increase in disc sizes is reflected in an increasing fragmentation fraction:
We find that in run ``5x~MHD'', none of the systems fragment.
For run ``8x~MHD'' we find \SI{0.19}{\%} of systems fragment, while in run ``12x~MHD'' the fraction of fragmenting systems is \SI{3.5}{\%}.
These values are computed by weighting the different mass bins such that the final stellar masses are consistent with the IMF.

Taken at face value, this puts the observed disc radii right at the edge of fragmentation, with only about 2 out of 1000 discs fragmenting. This would imply that a necessary condition for giant planet formation via gravitational instability is  only very rarely met (but also not never).  
However, the observed distribution of radii has a different, broader shape in comparison to our syntheses.

\subsection{Agreement with observed radii}\label{subs:obsrad}

\begin{figure}
  \includegraphics[width=\linewidth]{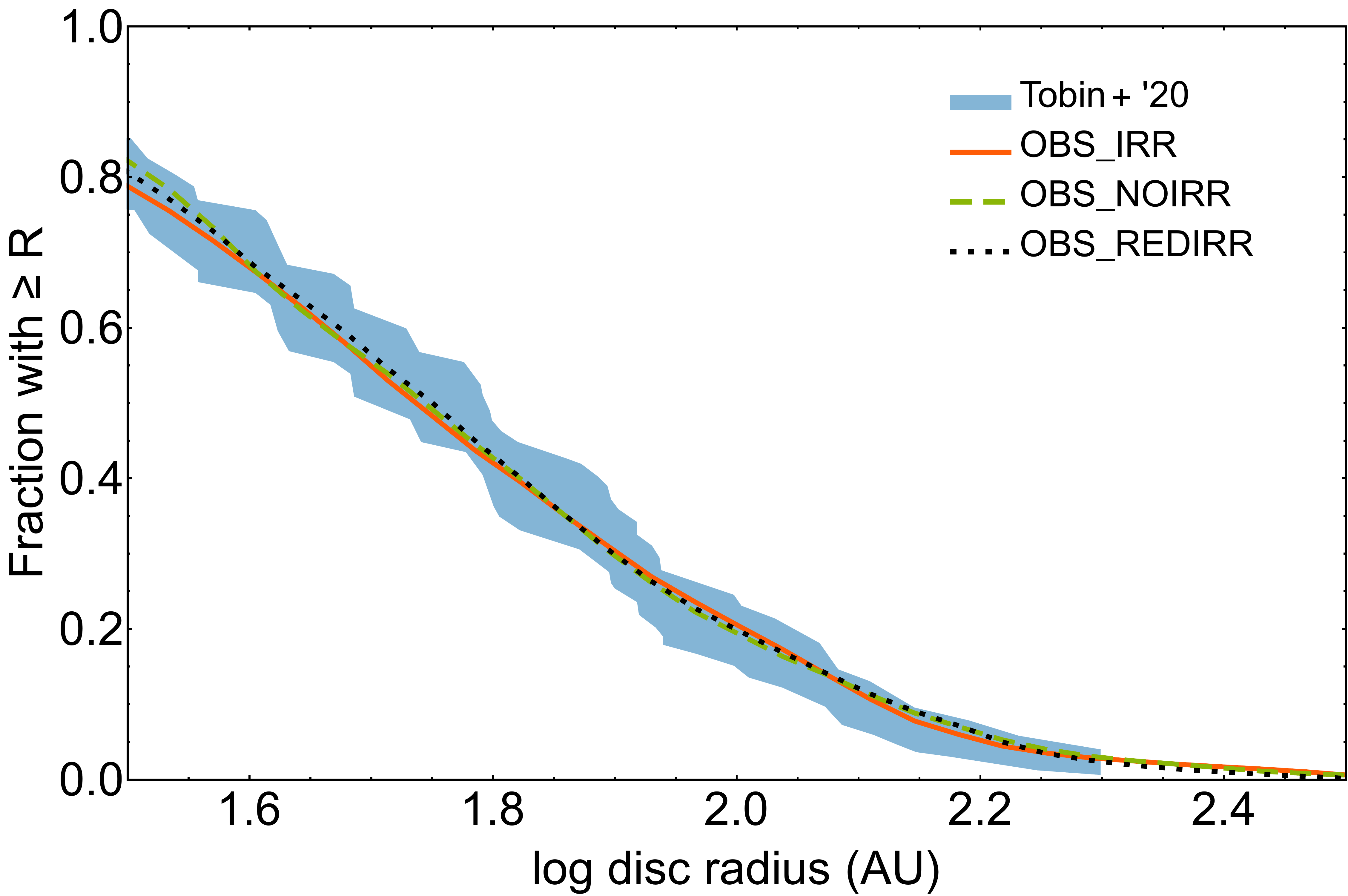}
  \caption{Cumulative distributions of the disc radii at the end of the infall phase for runs OBS\_IRR, OBS\_NOIRR and OBS\_REDIRR together with observed radii.}
  \label{fig_rad2}
\end{figure}

The different shapes of the size distributions discussed raise the question whether a synthetic population of discs with a distribution of radii in (better) agreement with observations can also be simulated.
This is not straight-forward, since there is no simple relationship between the infall radius and the early disc radius.
The disc's radius changes with time through the evolution of the disc and this effect depends on the disc's mass.
In Sect.~\ref{sect:Model} we describe how our simulations are distributed in different mass bins.
More massive systems tend to produce larger discs in our model. Such a correlation is also seen in the hydrodynamic simulations of \citet{2018MNRAS.475.5618B} as well as in observed discs \citep{2020ApJ...890..130T}. It is also expected from Eq.~\ref{eq_rhen}.
Increasing the infall radii in the bins corresponding to higher masses would therefore lead to a broader distribution of radii.
Of course, for such a modified set of initial conditions, the infall times need to be adapted again in order to satisfy the requirement to fit the IMF (Sect.~\ref{sect:Model}) which would affect the distribution of radii. 
After iterations, we infer the distribution of infall radii that satisfies both constraints at the same time (run OBS\_IRR).
The resulting distribution of radii is shown in Fig.~\ref{fig_rad2} (orange solid line).
The fraction of discs fragmenting in run OBS\_IRR is \SI{0.12}{\%}, similar to run ``8x~MHD''. %
This is expected as these two runs have a very similar distribution of radii above \SI{80}{au}, where fragmentation is most likely.
If this result is robust, it means fragmentation is very rare indeed, independently of subtleties in the disc radius distribution. 

\subsection{Irradiation from stellar accretion heating}\label{subs:irr}

Next we revisit our assumption concerning stellar accretion (shock) heating.
If this effect is indeed overestimated by our treatment as mentionned in Sect.~\ref{sect:Model} and discussed below, this would lead to too high temperatures in the outer discs and inhibit fragmentation through Eq.~\ref{eq:toomre}.
We tested this hypothesis by setting as a limiting case the second summand in Eq.~\ref{eq_lstar} to zero ($f_{\rm acc}$=0), effectively reducing the irradiation to the star's intrinsic luminosity.
We then constructed a new set of initial conditions that again satisfies the constraints in stellar mass and disc radius: run OBS\_NOIRR.
This run is also shown in Fig.~\ref{fig_rad2}.
In run OBS\_NOIRR, a high fraction of \SI{25}{\%} of the systems fragment, demonstrating the strong influence of accretion heating on fragmentation besides disc size. 
The assumptions about accretion heating made in runs OBS\_IRR ($f_{\rm acc}$=1) and OBS\_NOIRR ($f_{\rm acc}$=0) are likely extremes. The former case appears, however, clearly more likely than the latter, because there are direct observations \Citep{2016Natur.535..258C}
that the high stellar luminosity during episodes of strong accretion shifts the water iceline to large orbital distance, implying that $f_{\rm acc}$ must be larger than zero. We discuss this in more detail in Sect.~\ref{sect:Discussion}. 

\begin{figure}
  \includegraphics[width=\linewidth]{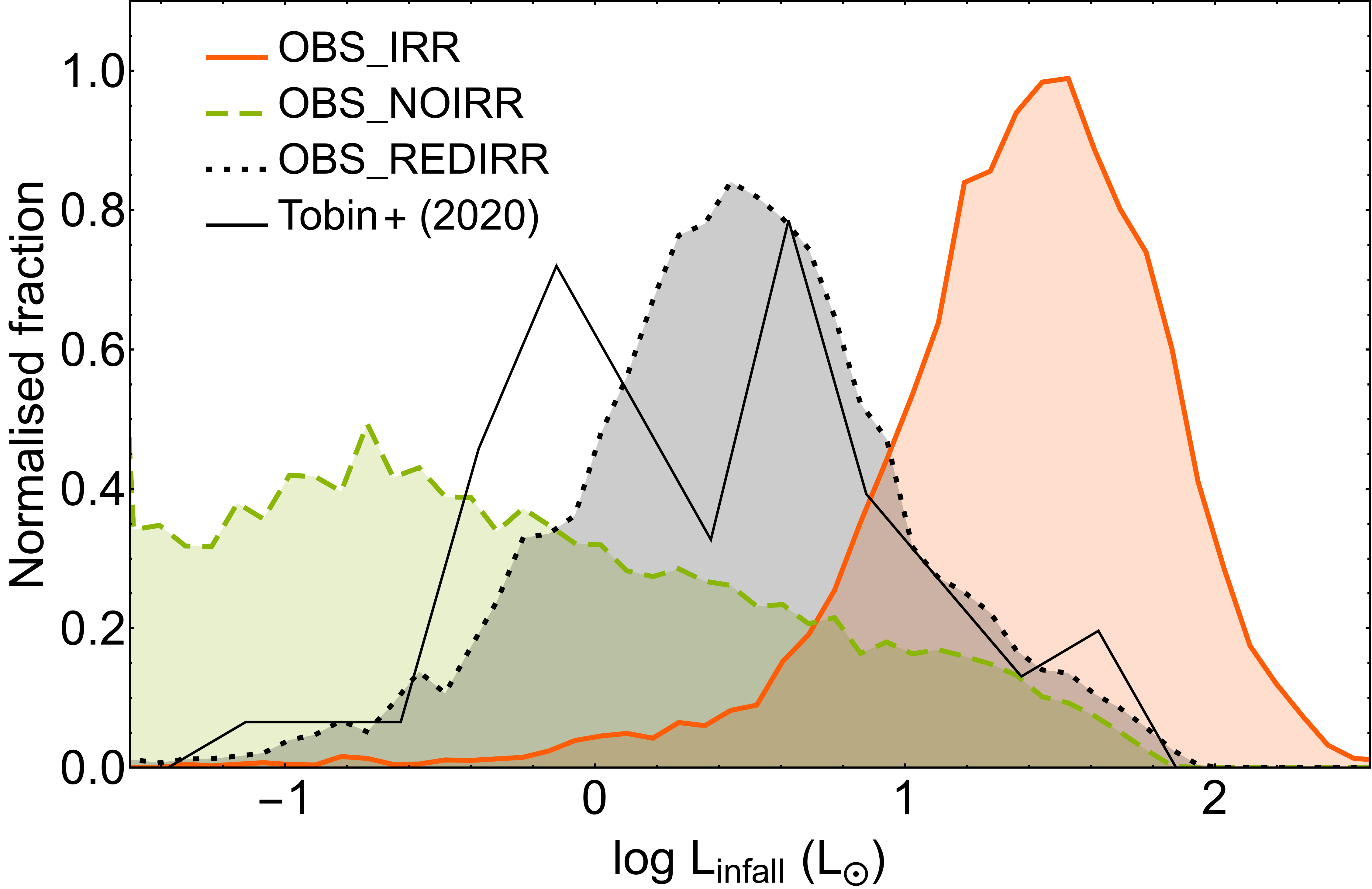}
  \caption{Total luminosities (intrinsic and accretion) at the end of the infall phase for runs OBS\_IRR, OBS\_NOIRR and OBS\_REDIRR. The black solid line shows observed luminosities of isolated Class~0 protostars from \citet{2020ApJ...890..130T}.}
  \label{fig_lum}
\end{figure}

Fig.~\ref{fig_lum} shows the stellar luminosities at the end of the infall phase together with the observed luminosities of isolated Class~0 protostars  \citep{2020ApJ...890..130T}.
The figure shows first that a very large range in luminosities is covered. Second, we note that the synthetic luminosity distributions when including only the intrinsic luminosity ($f_{\rm acc}$=0) or when considering the the sum of intrinsic and full accretion luminosity $f_{\rm acc}$=1) are bracketing the observed luminosity distribution. This suggest to consider an intermediate value for $f_{\rm acc}$. In run OBS\_REDIRR we thus assess a scenario with such reduced accretion heating.
To roughly reproduce the observed luminosity distribution, we find that the accretion term (second summand in Eq.~\ref{eq_lstar}) needs to be reduced by about a factor of 12, i.e., $f_{\rm acc}$=1/12. 
Again we chose infall radii and infall times in such a way that the constraints from both disc sizes and final stellar mass distribution are satisfied.
The fraction of systems that fragment is \SI{11}{\%} in run OBS\_REDIRR. 
This fragmentation fraction lies between what we found for runs OBS\_IRR ($f_{\rm acc}$=1, 0.12\% fragmentation) and OBS\_NOIRR ($f_{\rm acc}$=0, 25\% fragmentation) as expected: fragmentation is reduced less than in run OBS\_IRR because the discs are not as strongly heated and thus stabilised. The number is also interesting in another context: the frequency of substellar companions on wide orbits, which might have formed via gravitational instability, is on the order of a few to a few ten percent \citep{2021A&A...651A..72V}.

Even though our approach to fit the luminosity distribution is simplistic, run OBS\_REDIRR is still our most realistic, observationally-informed scenario. 
It satisfies the constraints from observed stellar masses and disc sizes while also exhibiting luminosities in the same range as is observed.
We discuss some limitations of our study in the following section.
In Appendix~\ref{app:prop} we show additional results for runs OBS\_IRR, OBS\_NOIRR and OBS\_REDIRR: early disc masses, final stellar masses and disc lifetimes of these systems.
The infall radii used in these runs are given in Appendix~\ref{app:radii}, the infall times in Appendix~\ref{app:tinfall}.

\subsection{Fragmentation}
\begin{figure*}[pt]
  \begin{subfigure}[pt]{0.49\textwidth}
  \includegraphics[width=\linewidth]{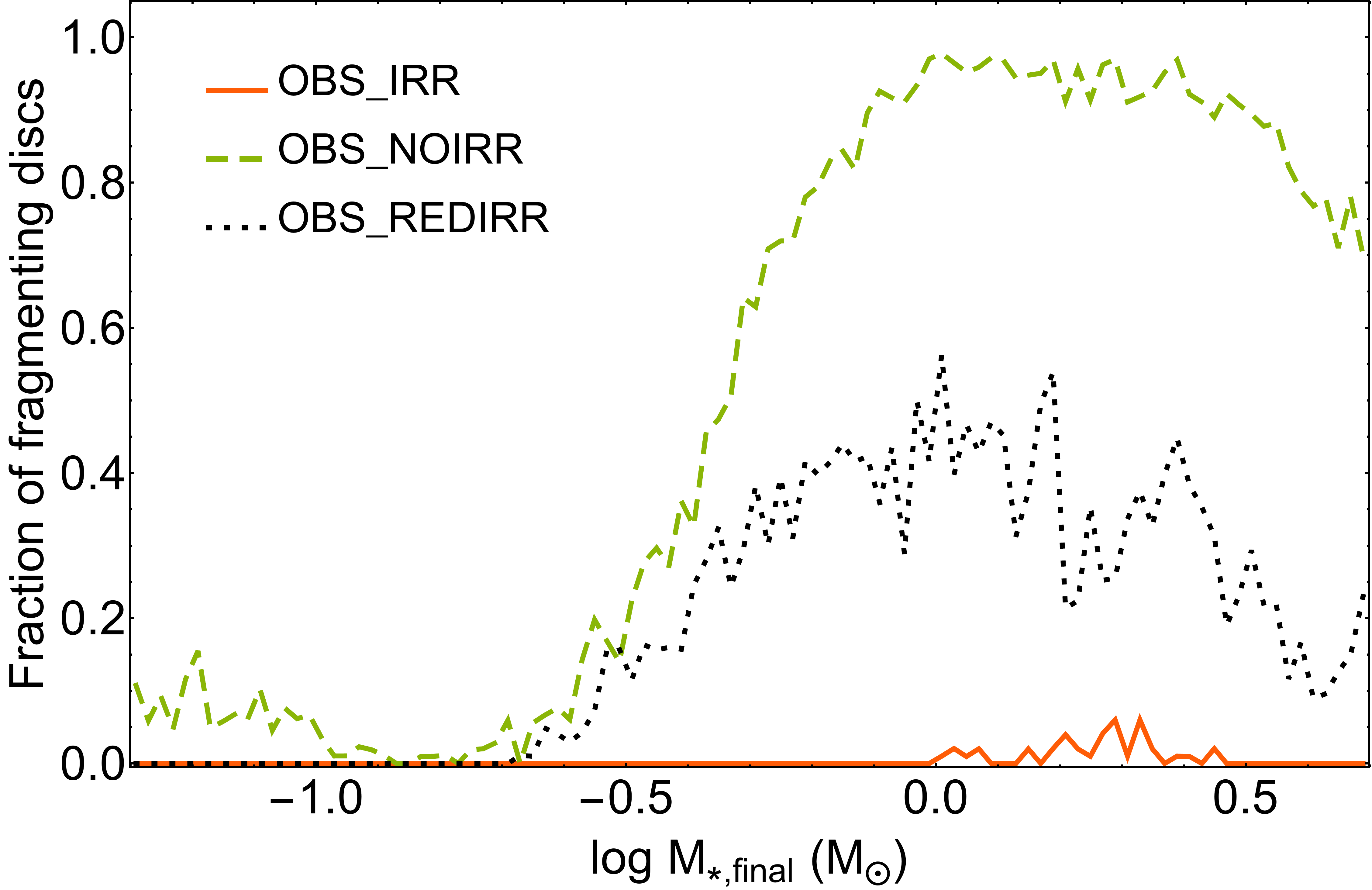}
  \end{subfigure}
  \begin{subfigure}[pt]{0.49\textwidth}
  \includegraphics[width=\linewidth]{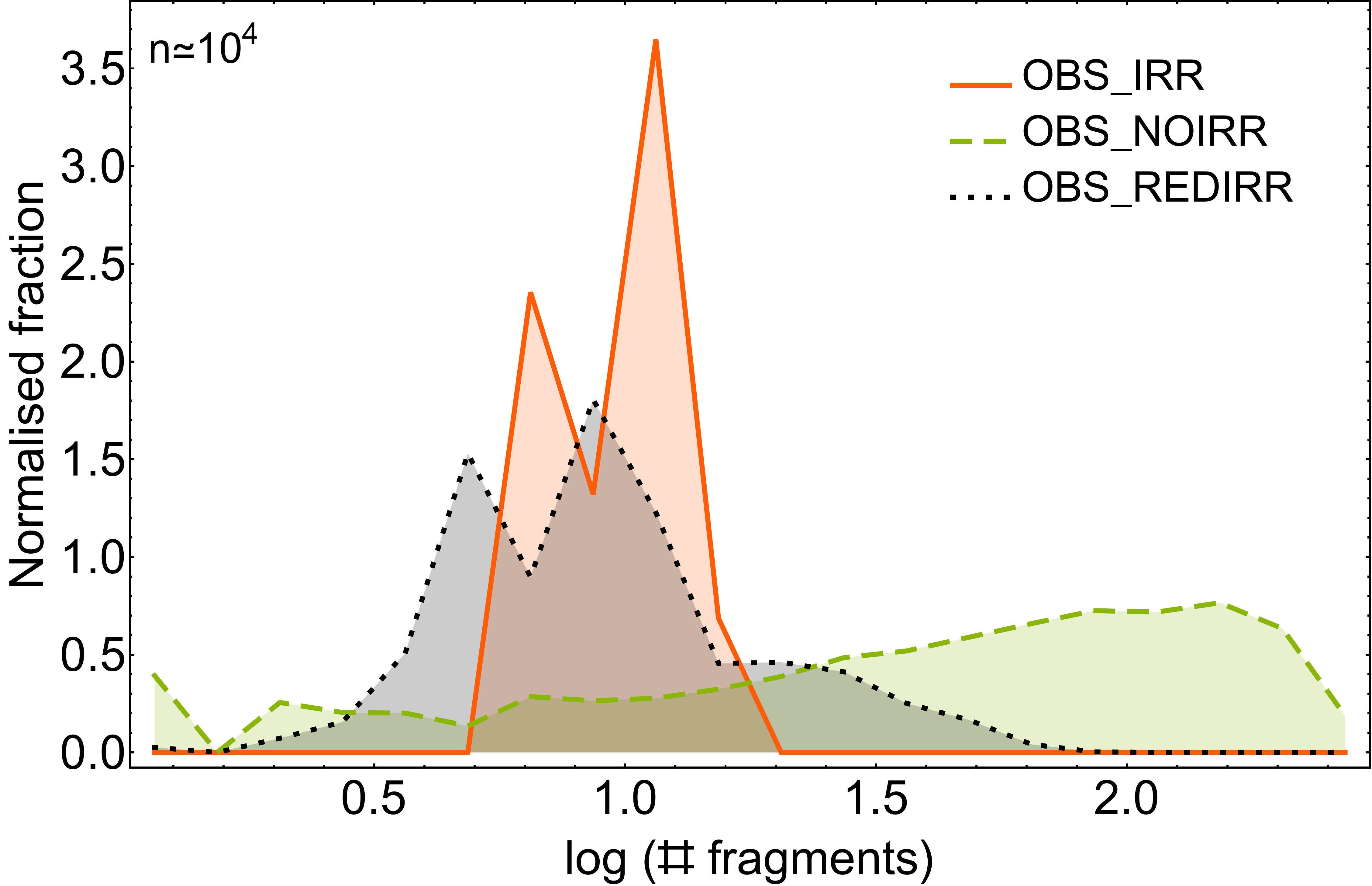}
  \end{subfigure}
  \begin{subfigure}[pt]{0.49\textwidth}
  \includegraphics[width=\linewidth]{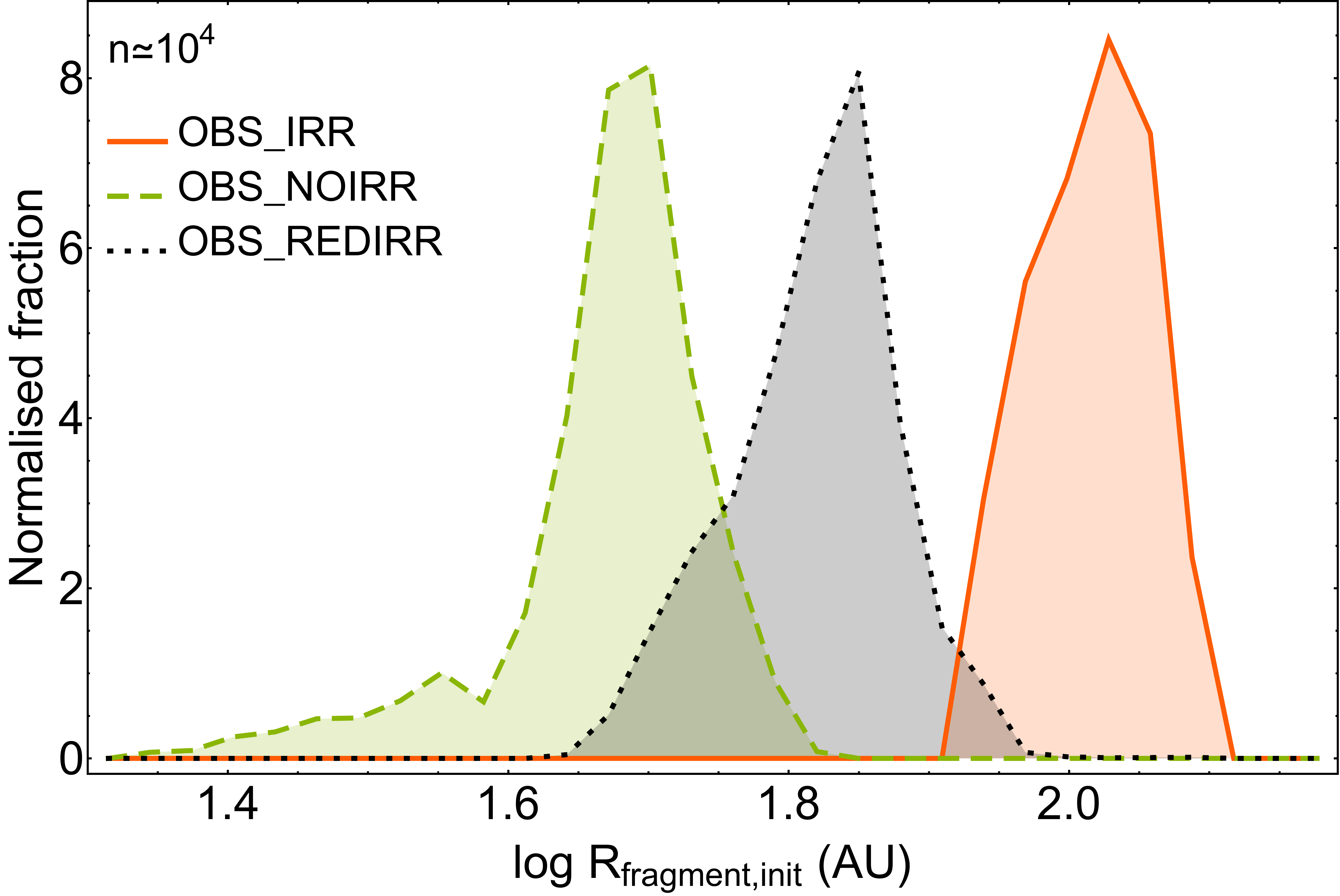}
  \end{subfigure}
  \begin{subfigure}[pt]{0.49\textwidth}
  \includegraphics[width=\linewidth]{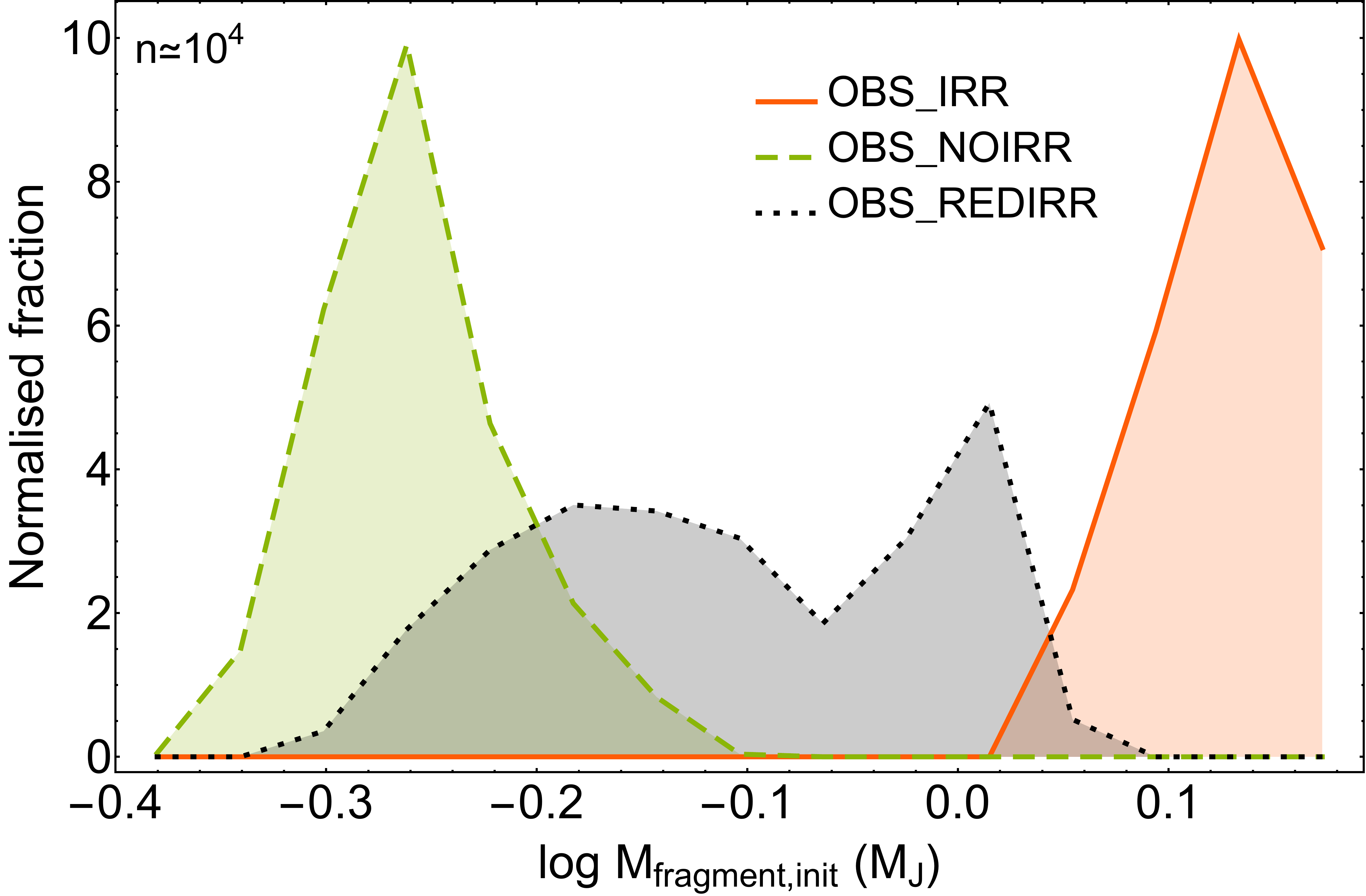}
  \end{subfigure}
  \caption{Results related to fragmentation for runs OBS\_IRR, OBS\_NOIRR and OBS\_REDIRR. Top left: Fraction of fragmenting discs as a function of final stellar mass. Top right: Mean number of fragments (for systems that do fragment). Bottom left: Location where the discs fragment. Bottom right: Initial fragment masses.}
  \label{fig:frag}
\end{figure*}

The likelihood of the disc to fragment and the location of the instability depend on the evolution of the disc's surface density and temperature.
Table~\ref{table_fragresults} gives an overview of the most important properties for our runs OBS\_NOIRR and OBS\_REDIRR.
\begin{table}
  \centering
  \begin{tabular}{ccccc}  
  \hline\hline
  \begin{tabular}[c]{@{}c@{}}Run\end{tabular} &
  \begin{tabular}[c]{@{}c@{}}Infall radius \end{tabular} &
  \begin{tabular}[c]{@{}c@{}}Mean \# of\\fragments\end{tabular} &
  \begin{tabular}[c]{@{}c@{}}$R_\mathrm{fr,init}$\\(\SI{}{au})\end{tabular} &
  \begin{tabular}[c]{@{}c@{}}$M_\mathrm{fr,init}$\\(\SI{}{M_J})\end{tabular} \\
  \hline
  1 & OBS\_IRR &  $\num{10(1)}$ & \num{100}  & $\num{1.3}$ \\
  2 & OBS\_NOIRR & $\num{29(14)}$ & \num{41} & $\num{0.53}$ \\
  3 & OBS\_REDIRR & $\num{11(5)}$ & \num{61(2)}  & $\num{0.71}$\\
  \hline
  \end{tabular}
  \caption[]{Fragment properties for runs OBS\_IRR, OBS\_NOIRR and OBS\_REDIRR. The columns show the mean number of fragments (for systems that do fragment) as well as the mean initial location ($R_\mathrm{fr,init}$) and the mean mass ($M_\mathrm{fr,init}$) of the fragments.}
  \label{table_fragresults}
\end{table}
Fig.~\ref{fig:frag} shows the fraction of fragmenting discs as a function of final stellar mass as well as the mean values of the number, location and mass of the fragments.
The top left panel reveals that fragmentation only becomes important for systems with a final stellar mass $\gtrsim \SI{0.3}{\msun}$, with typical values of $\sim \SI{40}{\%}$ in run OBS\_REDIRR and $\sim \SI{90}{\%}$ in run OBS\-NOIRR. The fraction decreases slightly for final stellar masses above \SI{3}{\msun}. This was not seen in \citetalias{2021A&A...645A..43S} and is a consequence of the different choice of infall radii. Interestingly, a similar trend is found in radial velocity surveys. An increase in giant planet occurrence rate with host star mass is found up to a host star mass of $\SI{1.9}{\msun}$ with a decline at higher host masses (e.g. \citealt{2015A&A...574A.116R}). It is still unclear whether the trend in fragmentation fractions persists in fully formed giant planets. Also, it is still unknown whether a similar decrease at higher stellar masses exists in directly imaged planets due to insufficient detections at higher stellar masses (\citealt{2021A&A...651A..72V}, see also \citealt{2021A&A...646A.164J,2022AJ....163...80W}).
The top right panel of Figl~\ref{fig:frag} shows that the number of fragments depends strongly on irradiation. While the runs with accretion heating of the outer disc (OBS\_IRR and OBS\_REDIRR) typically have around ten fragments, this number is larger without accretion heating and can lead to more than a hundred fragments.
The number of fragments in this case is lower than what we found in \citetalias{2021A&A...645A..43S}, less than half in the OBS\_IRR and OBS\_REDIRR runs compared to the ``hydro'' run. This is a consequence of the more compact discs.

The location where the discs fragment (bottom left panel in Fig.~\ref{fig:frag}) depends mainly on the accretion heating: The hotter the outer discs are, the further out they fragment, in agreement with \citetalias{2021A&A...645A..43S}.

The initial fragment mass (bottom right panel) is influenced both by the location of fragmentation and the accretion heating. More accretion heating means higher temperatures in the outer disc which leads to more massive fragments through Eq.~\ref{eq_mtoomre} directly. It also does so indirectly by moving the location of fragmentation further out. This is in good agreement with what was found in \citetalias{2021A&A...645A..43S}. Run OBS\_REDIRR features quite low initial fragment masses, typically \SIrange{0.5}{1}{M_J}.

\section{Discussion}\label{sect:Discussion}
In our analysis we had to  make several strong assumptions due to theoretical and observational uncertainties.
Here we discuss how these may influence our results.

The most important assumption is the comparability of gas and dust disc radii.
For older discs like Class~II this would certainly not be justified, as dust would have had time to grow and drift \Citep{2010A&A...513A..79B,2014prpl.conf..339T,2020ApJ...890..130T}.
Since we compare the sizes of young discs (Class~0), the differences are expected to be smaller. However, it is still possible that there is a mismatch between observed Class~0 dust discs sizes and the sizes of the corresponding gas discs in both directions.

The definition of the disc radius used in this work (\SI{63.2}{\%} of the mass) is identical to that used in \citet{2018MNRAS.475.5618B}.
It is well suited for comparisons with observations of discs with truncated power-law surface density profiles $\Sigma \propto r^{-\gamma}$ independent of the index $\gamma$ (as long as $\gamma<2$, see the discussion in Sect.~2.3.1 in \citealt{2018MNRAS.475.5618B}). 
For our comparison we evaluate the disc radii at a well-defined point in time: the end of the infall phase.
If this coincides with the end of the Class~0 phase, this means we overestimate the disc sizes, since the observed discs are expected to to represent an (unknown) distribution of Class~0 ages.
If on the contrary, the Class~0 phase lasts longer than our infall phase and the discs would have more time to spread, or if the size of the observed gas disc is smaller due to growth/drift of dust particles this would make our discs too small.
Based on our discussion in Sect.~\ref{sect:result}, an overestimate of the disc sizes would lead to an overestimate of the fraction of discs that fragment.
Conversely if our discs are too small, we are likely to underestimate the occurrence rate of fragmentation.

Our results indicate that only a minority of discs undergo fragmentation.
This does not mean, however, that gravitational instabilities are not important in the non-fragmenting bulk of the population.
In fact, most of the discs in run OBS\_REDIRR have $Q_\mathrm{Toomre} < 2$ somewhere, at some point during their evolution. This suggests  that these discs were self-gravitating for at least a short period during their early lives.

In comparing the size distribution of our full sample with the sample from \citet{2020ApJ...890..130T} we also implicitly assume their sample will be representative of the IMF.
Unfortunately, it is very difficult to measure stellar masses of such young systems, and the precise distribution of stellar masses of the observed sample is unknown.
The observed luminosities cannot be translated into masses directly due to accretion.
Our comparison of luminosities with and without accretion (Fig.~\ref{fig_lum}) demonstrate the potential extent of this discrepancy: the total luminosities in our run OBS\_IRR are higher by more than an order of magnitude compared to the observations, which are again much higher than our intrinsic luminosities.
This could be explained in part by a different distribution of masses in the observed sample. Our populations have a stellar mass distribution as in the \citet{2005ASSL..327...41C}-IMF. However, a different IMF dominated may be more adequate for the Orion region (\citealt{2012ApJ...748...14D}, see also \citealt{2000ApJ...544.1044L}).
The difference could also be caused  by the way the stars accrete.
The difference between observed and theoretically expected luminosities has long been discussed and is historically known as the ``protostellar luminosity problem'' \Citep{1990AJ.....99..869K}.
This apparent problem has been solved by demonstrating that both episodic accretion and a longer accretion phase can produce luminosities that are in line with observations (e.g. \citealt{2011ApJ...736...53O,2012ApJ...747...52D}).

The precise nature in which accretion happens is still not fully understood. Also, it is still unknown which of the two proposed mechanisms described above contributes to what degree to the explanation of the ``protostellar luminosity spread'' (Sect.~3 in \Citealt{2022arXiv220311257F}).
It is therefore unclear, how our assumption of a phase of constant accretion should be modified in order to account for the observed luminosity spread. This is an important topic that needs to be studied further. The exact nature of accretion affects the stellar evolution, and hence disc fragmentation. A study that assesses the interplay of episodic accretion, and stellar evolution in a 3D MHD simulation is presented by \citet{2018MNRAS.475.2642K}.

Finally, we also assumed that once the disc fragments, a clump of a given mass $M_F$ (Eq.~\ref{eq_mtoomre}) forms, migrates inwards quickly and accretes on the star. 
While this work does not  focus on the fate of the clumps, it should be noted that our simplifying assumptions  could influence the disc's evolution and its fragmentation at later stages.
While $M_F$ is the initial clump's mass, it could be significantly different  from the mass accreted onto  the star and/or the final mass of  the  surviving clump.
The clump's mass is expected to decrease via mass loss (e.g., tidal disruption in the inner disc, \citealt{2010Icar..207..509B,2010MNRAS.408L..36N}). 
The forming clump could also grow in mass via further accretion of disc gas bringing it into the Brown Dwarf or stellar mass regime. If gas accretion is rapid, a gap may form which would slow down migration \citep{2008ApJ...685..560D,2020MNRAS.496.1598R,2012ApJ...746..110Z,2020A&A...644A..41O,2022A&A...664A.138S}. A deep gap could reduce gas flow to the inner disc and make the outer disc prone to further fragmentation.

\section{Summary, conclusions, and outlook}\label{sect:Conclusions}

\begin{figure}
  \includegraphics[width=\linewidth]{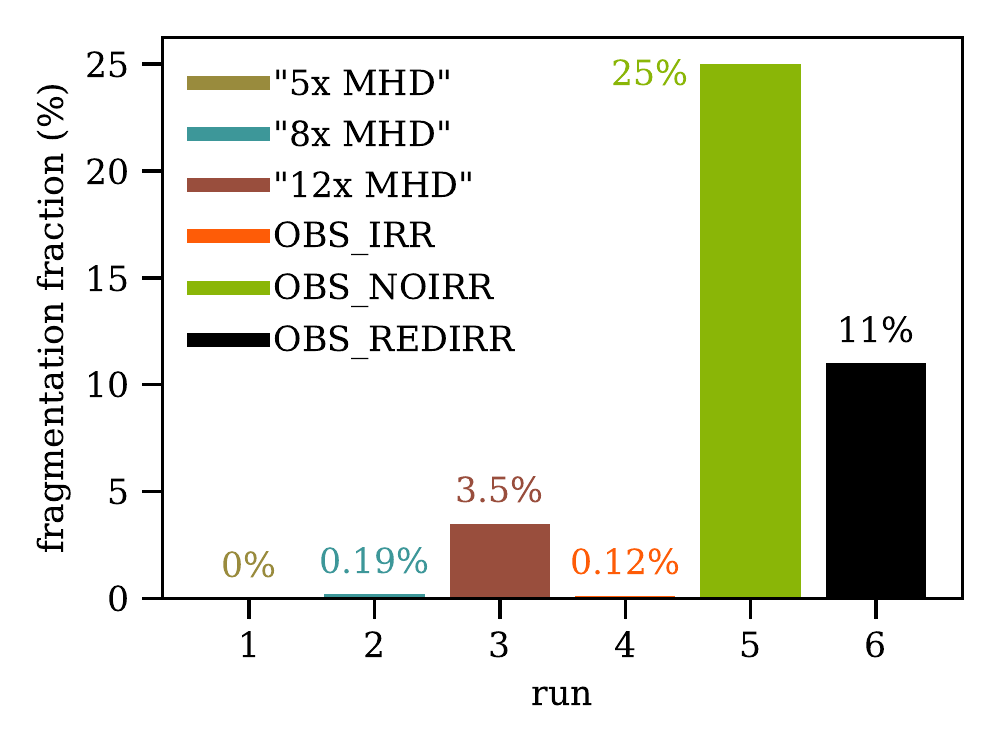}
  \caption{Fraction of discs that fragment for all the runs studied. Each run consists of \num{10000} simulations and the results are representative of a distribution of final masses in agreement with the IMF. The most realistic scenario is run OBS\_REDIRR, where the stellar shock luminosity is scaled to agreed with observed luminosities of Class 0 protostars.}
  \label{fig_res}
\end{figure}

We performed a population synthesis of protoplanetary discs focusing on disc sizes and the fragmentation likelihood. 
In three sets of initial simulations (runs ``5x~MHD'', ``8x~MHD'' and ``12x~MHD'') we demonstrated how increasing the infall radii controls the fragmentation of protoplanetary discs.
We then constructed initial conditions (infall locations) in such a way that the synthetic disc radius distribution at the end of the infall phase agrees with the observed radius distribution of Class~0 discs, while the distribution of stellar masses at the end of our simulations agrees with the IMF (runs OBS\_IRR, OBS\_NOIRR and OBS\_REDIRR).
For these three runs, we also investigated how fragmentation is influenced by accretion heating. Our most important results are summarised in Fig.~ \ref{fig_res}.
Our key conclusions can be summarized as follows:
\begin{itemize}
	\item Protoplanetary discs need to reach a certain size while they are massive in order to fragment. Observed discs appear to have radii just large enough to fragment in a few cases.
	\item  The early disc size distribution is mainly determined by the infall location. To get a synthetic size distribution that is in agreement with the observed one, we find infall locations that are between those predicted by some magnetized and pure hydrodynamic star formation simulations. Specifically, the infall locations need to be increased by factors 5-8 relative to the MHD simulations of \citet{Hennebelle2016} which corresponds to a reduction by factors 2-3 relative to pure HD simulations \citetalias{2018MNRAS.475.5618B}.
	\item The fraction of discs that actually fragments does, however, not only  depend on their early sizes, but also on the fraction $f_{\rm acc}$ with which stellar accretional (shock) luminosity contributes to heating the outer disc.
	\item In disc population syntheses fitting the observed Class 0 disc radius distribution, we find fragmentation fractions of 0.12\% and 25\% for the limiting cases of  $f_{\rm acc}$=1 and 0, respectively. Here, the former scenario is more likely than the latter, since heating of discs by stellar accretion luminosity is an observed phenomenon \citet{2016Natur.535..258C}.
	\item Finally, in our most realistic case, we empirically determine $f_{\rm acc}\approx$1/12 by requiring that our synthetic luminosity distribution is in approximate agreement with the observed luminosity distribution of Class 0 protostars (while still also fitting the observed disc radius distribution). In this case, the fragmentation fraction is 11\%, which is our observationally-informed best estimate of the fraction.  
	\item A better understanding of accretion is crucial for the understanding of the evolution of protoplanetary discs and the formation of planets. It needs to be investigated further using simulations of discs with an evolving dust component, including radiative transport, and evolution models of accreting stars.
\end{itemize}

The discovery of AB~Aurigae~b \citep{2022NatAs.tmp...76C} as a massive, still actively accreting giant planet at a large orbital distance of $\SI{93}{au}$ has recently given observational support to gravitational instability as a giant planet formation mechanism. However, it is still  unknown at which frequency GI contributes  to giant planet formation. Qualitatively, it is often assumed to be a rather rare process \Citep{2021A&A...651A..72V}, but quantitative theoretical predictions are currently sparse.
  
A necessary condition for giant planet formation is disc fragmentation. The fragmentation fractions of about 0.1 to 11\% that we find in this work are thus an upper bound for the frequency of giant planet formation via GI. Our work represents a step towards a quantitative theoretical prediction of the importance of GI as a giant planet formation mechanism. 

Future work promises to help in our understanding of the formation and evolution of stars, discs, and planets.
Simulations of star formation are becoming more and more detailed, allowing for a more precise treatment of the interaction between stars and discs \citep{2018MNRAS.474.1176J,2018A&A...616A.101K}.
The study of disc kinematics continues to improve our understanding of the mass distribution and motion of the gas in the outer discs and the presence of structures and embedded protoplanets \citep{2019Natur.574..378T,2020ApJ...890L...9P,2022MNRAS.510.1671T}.
Observations of the gas emission of discs may help constraining young disc sizes in the near future (e.g. \citealt{2022A&A...662A.121R}). Finally, exoplanet surveys continue to improve our  understanding of the demographics of planets, including in particular distant and forming planets which are prime candidates for a formation via GI \Citep{2008Sci...322.1348M,2017A&A...605L...9C,2022NatAs.tmp...76C}.
This will eventually allow to understand how star formation, infall, disc morphology, giant planet formation via gravitational instability, and the observed extrasolar planet population are connected. 

\begin{acknowledgements}
We thank the anonymous referee for valuable comments.
We also thank Til Birnstiel, Rolf Kuiper and Lucio Mayer for the insightful discussions. The authors acknowledge the financial support of the SNSF. O.S. and C.M. acknowledge the support from the Swiss National Science Foundation under grant \texttt{\detokenize{200021_204847}} ``PlanetsInTime''.
R.H. acknowledges support from SNSF grant \texttt{\detokenize{200020_188460}}. Parts of this work has been carried out within the framework of the NCCR PlanetS supported by the Swiss National Science Foundation under grants \texttt{\detokenize{51NF40_182901}} and \texttt{\detokenize{51NF40_205606}}.
\end{acknowledgements}

\section*{ORCID iDs}
Oliver\,Schib\,\orcidicon{}\\ \url{https://orcid.org/0000-0001-6693-7910}\\
Christoph\,Mordasini\,\orcidicon{}\\ \url{https://orcid.org/0000-0002-1013-2811}\\
Ravit\,Helled\,\orcidicon{}\\ \url{https://orcid.org/0000-0001-5555-2652}\\

\bibliographystyle{aa}
\bibliography{library.bib}

\clearpage

\begin{appendix}
\section{Properties of runs OBS\_IRR, OBS\_NOIRR and OBS\_REDIRR}\label{app:prop}
\begin{figure}
  \includegraphics[width=\linewidth]{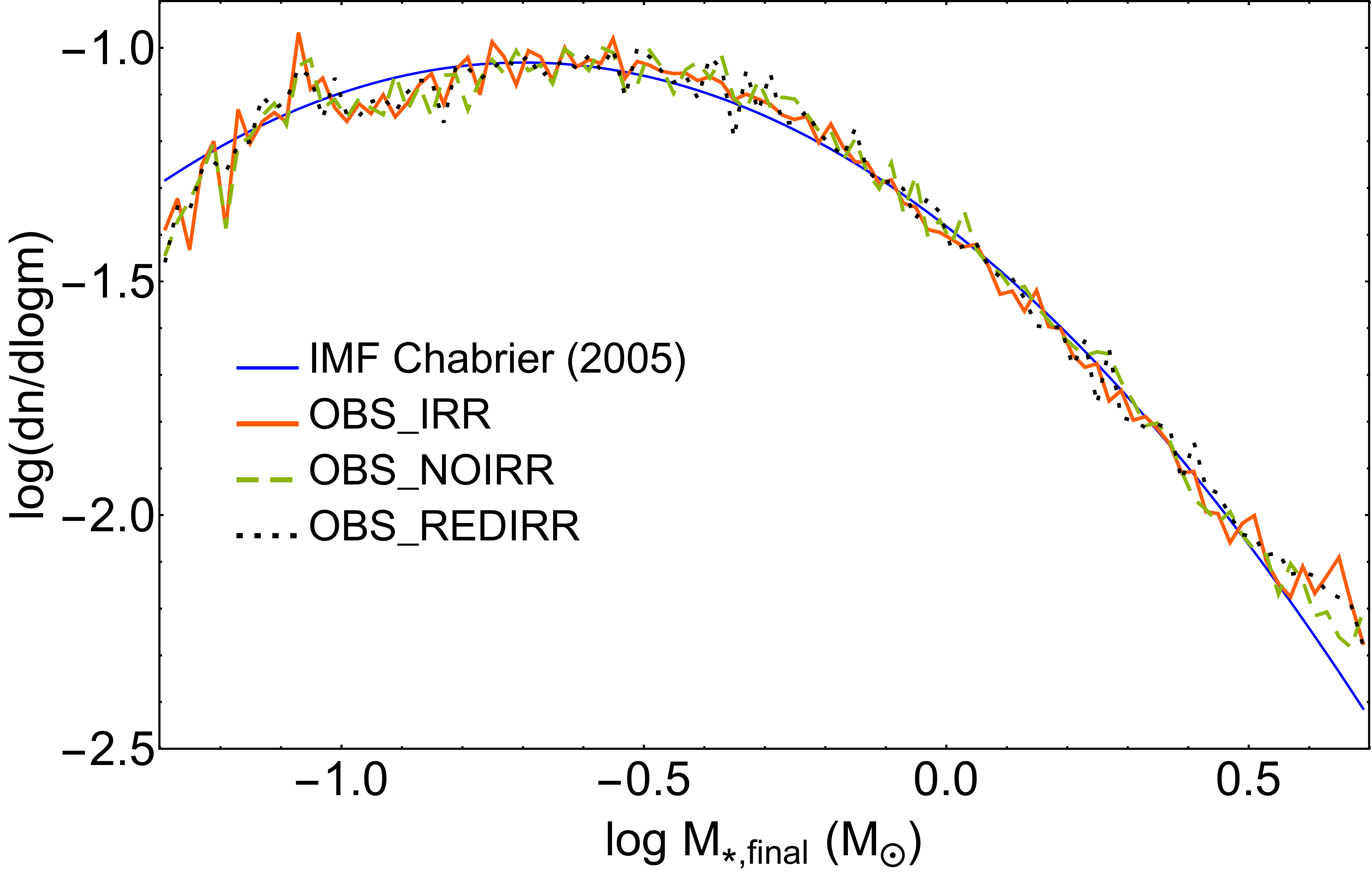}
  \caption{Stellar masses at the end of the simulation for runs OBS\_IRR, OBS\_NOIRR and OBS\_REDIRR, compared to the \citet{2005ASSL..327...41C} IMF.}
  \label{fig_imf}
\end{figure}

\begin{figure}
  \includegraphics[width=\linewidth]{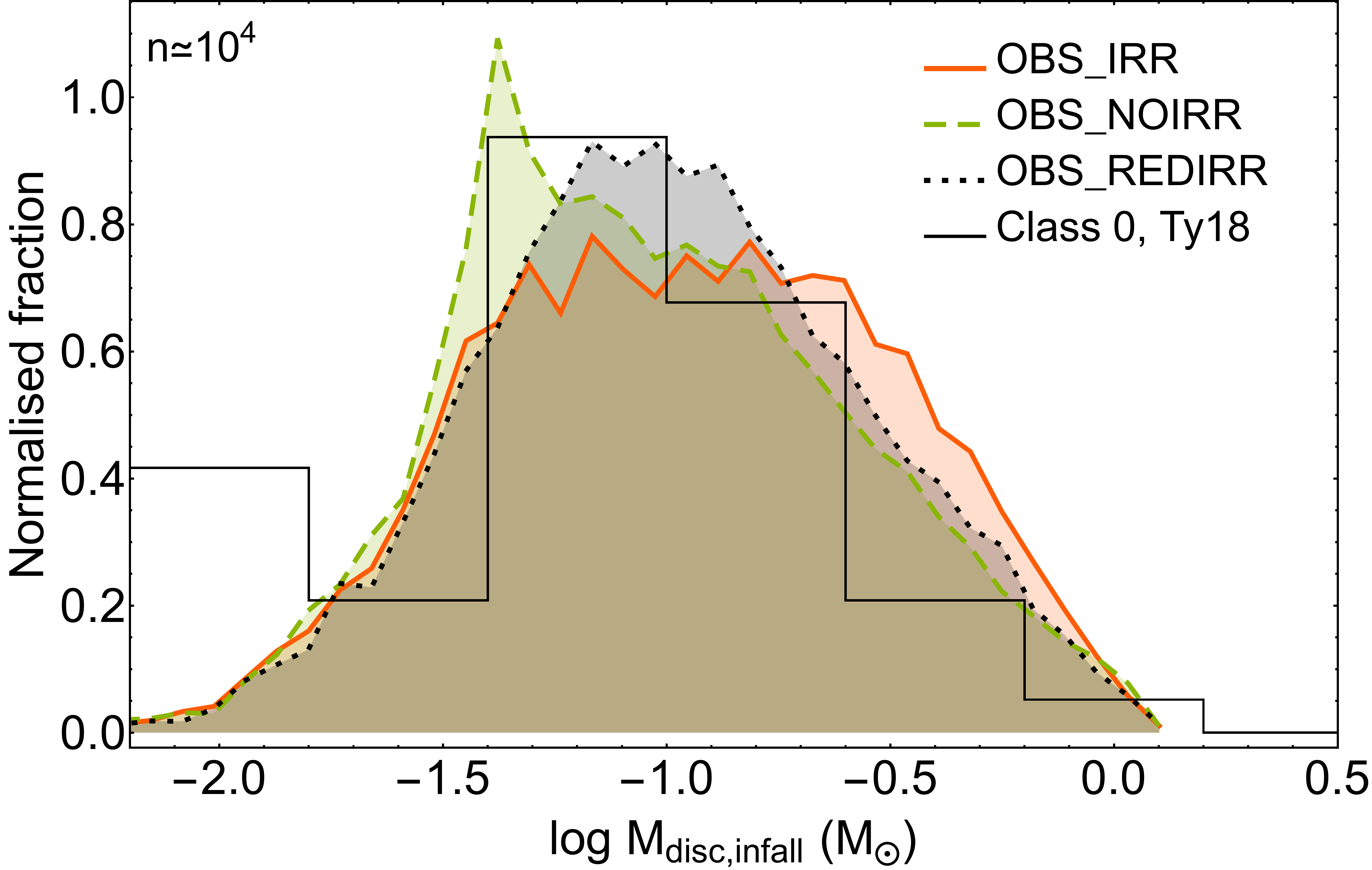}
  \caption{Disc masses at the end of the infall phase for runs OBS\_IRR, OBS\_NOIRR and OBS\_REDIRR, compared to observed Class~0 discs.}
  \label{fig_mdisc}
\end{figure}

\begin{figure}
  \includegraphics[width=\linewidth]{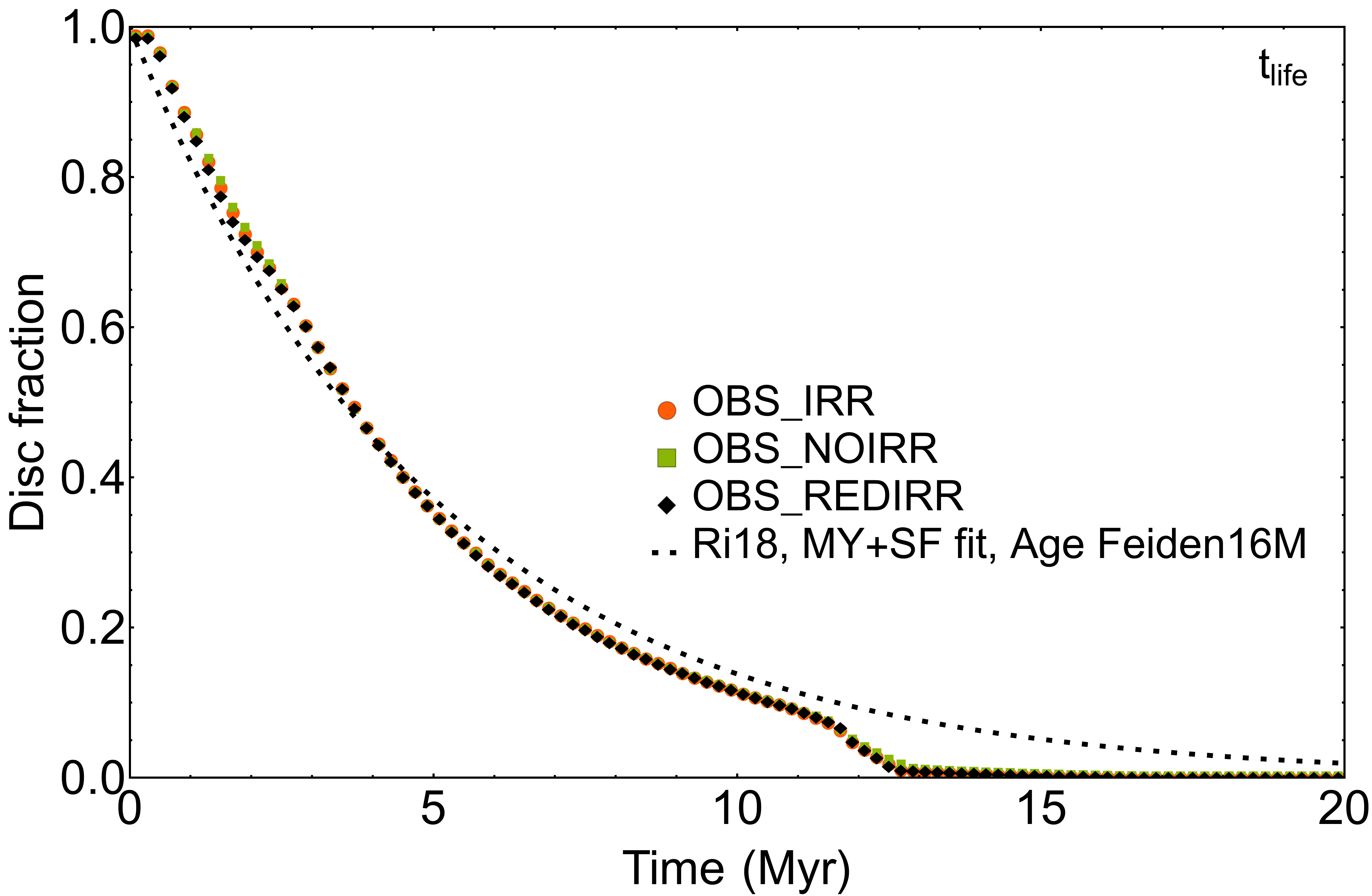}
  \caption{Disc fractions based on reduced disc lifetimes as a function of time for runs OBS\_IRR, OBS\_NOIRR and OBS\_REDIRR compared to a fit to observed disc fractions \citep{Richert2018}.}
  \label{fig_tlife}
\end{figure}

Here we show some additional results for runsOBS\_IRR, OBS\_NOIRR and OBS\_REDIRR.
This can be useful when comparing these runs to the ones from \citetalias{2021A&A...645A..43S} or other studies.
An overview of these results is given in Table~\ref{table_results}.
Fig.~\ref{fig_imf} depicts the distributions of stellar masses at the end of the simulations (i.e. when the disc is gone).
It shows that the agreement with the IMF is reasonable for all runs.
In Fig.~\ref{fig_mdisc} we show the distribution of disc masses at the end of the infall phase.
Also shown is the observational result from \citet{2018ApJS..238...19T}.
The masses are similar for runs OBS\_IRR, OBS\_NOIRR and OBS\_REDIRR. They are comparable, tough somewhat larger than the observed Class~0 masses.
Systems with less accretion heating are slightly less massive.
This result may seem counter-intuitive, since lower disc temperatures should lead to less accretion to the star and therfore more massive discs.
However the infall radii in the less massive (and more abundant) systems are lower for systems with less accretion heating. 
herefore more gas is accreted on the star early, which explains the lower masses.
Figure~\ref{fig_tlife} depicts the disc fractions based on the reduced near-infrared lifetimes $t_\mathrm{life}$.
It shows excellent agreement with the disc fractions based on the fits from \citet{Richert2018}, when the ``magnetic'' pre main sequence (PMS) model from \citet{Feiden2016} is used to determine cluster ages.
The reduction is based on the beginning of PMS \citep{2016MNRAS.461.2257K}.
We note that there is substantial uncertainty involved when comparing disc lifetimes from simulations to observed disc fractions.
A detailed discussion of this is found in Sect.~6.3 of \citetalias{2021A&A...645A..43S}.

\section{Infall radii}\label{app:radii}
\begin{figure}
  \includegraphics[width=\linewidth]{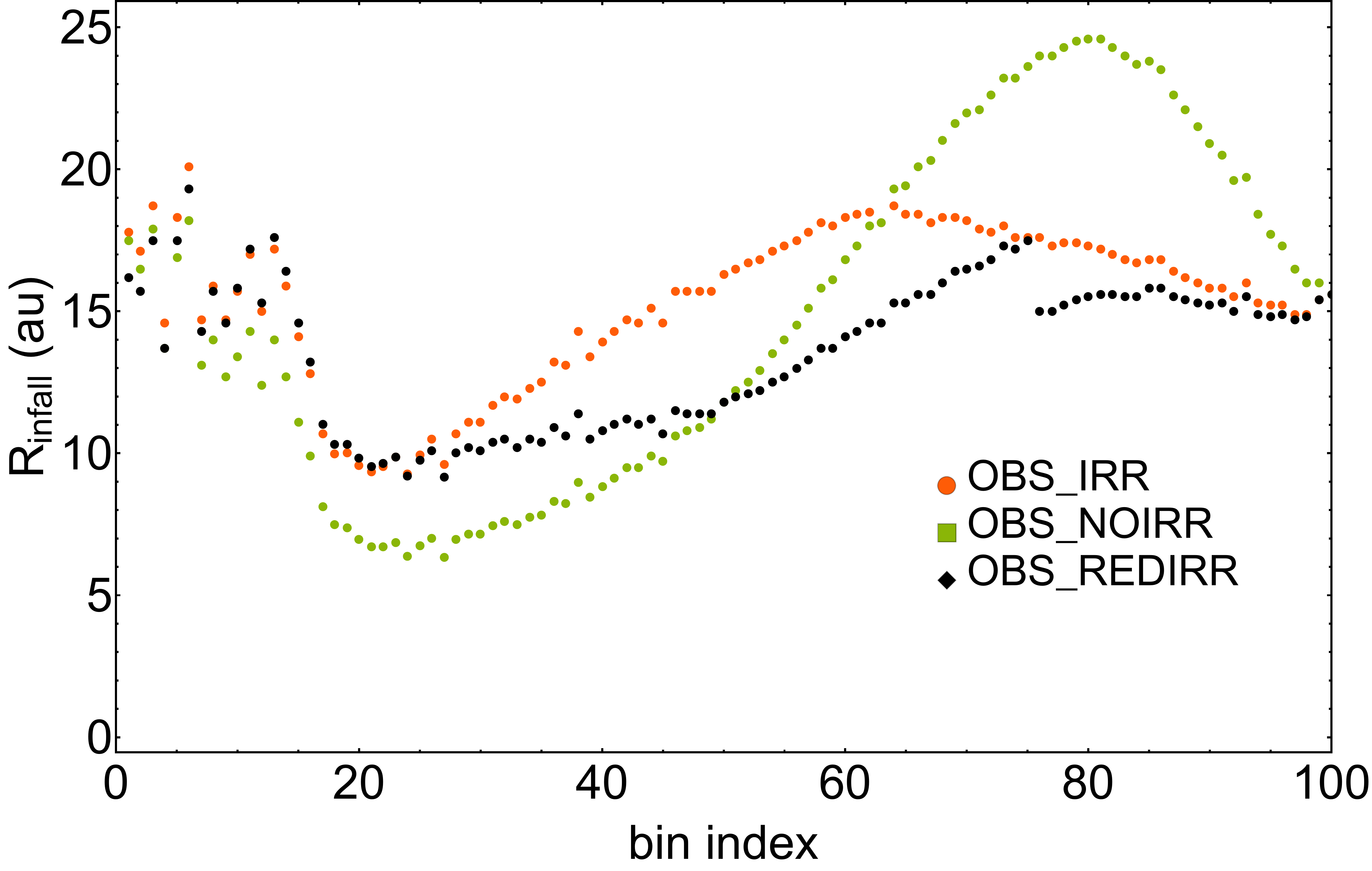}
  \caption{Infall radii for OBS\_IRR, OBS\_NOIRR and OBS\_REDIRR.}
  \label{fig_rinf}
\end{figure}

In Sect.~\ref{sect:Model} we describe how we vary the infall radii (the locations, where the infalling material from the MCC is deposited in the disc).
Figure~\ref{fig_rinf} displays the radii chosen for runs OBS\_IRR, OBS\_NOIRR and OBS\_REDIRR for reference.
A reduction of the accretion heating means the infall radii need to be reduced in bins with lower index (corresponding to smaller, less massive systems) and increased in the systems with higher index.
The definitions of the bins is given in Table~D.1 of \citetalias{2021A&A...645A..43S}.

\section{Infall times}\label{app:tinfall}
\begin{figure}
  \includegraphics[width=\linewidth]{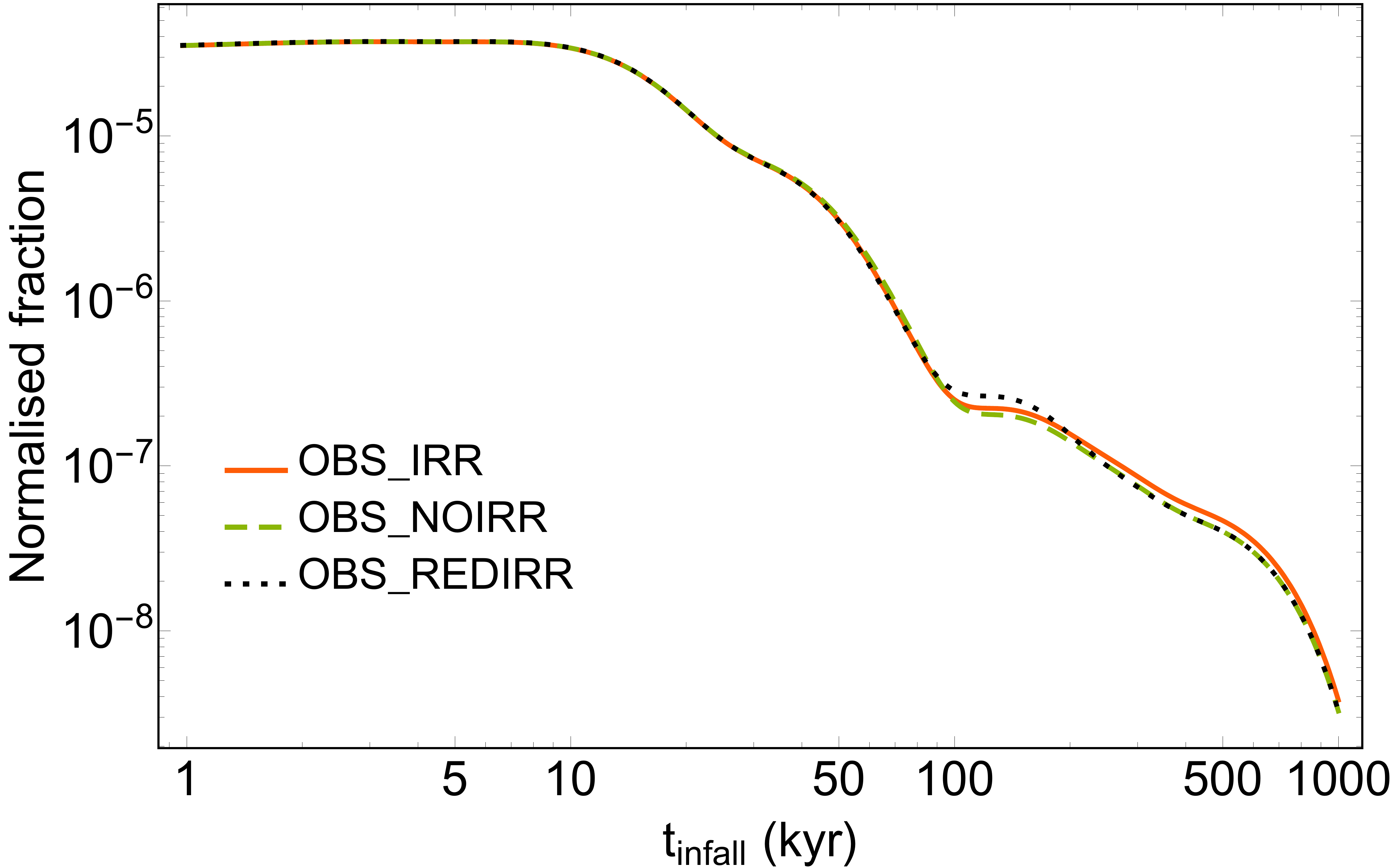}
  \caption{Infall times for runs OBS\_IRR, OBS\_NOIRR and OBS\_REDIRR.}
  \label{fig_tinf}
\end{figure}

The infall times (duration of the infall phase) need to be varied slightly from run to run in order for the finall stellar masses to agree with the IMF.
The distributions used in runs OBS\_IRR, OBS\_NOIRR and OBS\_REDIRR are shown in Fig.~\ref{fig_tinf}.
They should be compared to the bottom right panel of Fig.~E.1 in \citetalias{2021A&A...645A..43S}.
The distributions shown in Fig.~\ref{fig_tinf} are kernel density estimates from which the infall time for each system is drawn.
The same is done for the initial values for the stellar mass, disc mass and infall rate.
In contrast to the infall times, the distributions for the other initial values are identical to what is used in \citetalias{2021A&A...645A..43S} and are given there (Appendix~E).

\end{appendix}

\end{document}